\documentstyle[aps,epsf]{revtex}

\newcommand{\am}{a_{\mu}}
\newcommand{\az}{\alpha(M_{Z}^{2})}
\newcommand{\azinv}{\alpha(M_{Z}^{2})^{-1}}
\newcommand{\ee}{e^{+}e^{-}}
\newcommand{\jp}{J/\psi}
\newcommand{\mumu}{\mu^{+}\mu^{-}}
\newcommand{\ra}{\rightarrow}

\global\firstfigfalse  
\global\firsttabfalse  

\begin{document}

\baselineskip 14pt
\title{Measurement of R Between 2-5 GeV}
\author{Derrick Kong}
\address{University of Hawaii}
\maketitle

\begin{abstract}

We have obtained measurements of the total cross section for
$e^{+}e^{-}$ annihilation into hadronic final states for 6 energy points
(2.6, 3.2, 2.4, 3.55, 4.6, and 5.0 GeV) with the upgraded Beijing
Spectrometer (BESII).  We report preliminary values from this data and
outline future plans for a finer scan in the 2-5 GeV energy range.

\end{abstract}

\section{Introduction}

Two topics that depend on measurements in the center of mass (CM) energy
range of a few GeV are of current interest.  One is the precision of the
QED coupling constant evaluated at the mass of the $Z$ boson, $\az$, and
the other is the anomalous magnetic moment of the muon, $\am$.  Both are
accessible via a measurement of the lowest order cross-section for $\ee
\ra \gamma^{\ast}$ hadrons, which is usually parameterized in terms of
the ratio $R$, defined as

\begin{equation}
R \equiv \frac{\sigma(\ee \ra \mbox{hadrons})}{\sigma(\ee \ra \mumu)},
\end{equation}

\noindent
where the denominator is the lowest-order QED cross section and equals
$4\pi \alpha^{2} / 3s$.

The $R$ ratio has been measured by many experiments over a large CM
energy range from the hadron production threshold to the $Z^0$ mass
\cite{1}.  Experimentally determined $R$ values are, in general,
consistent with theoretical predictions, and provide an impressive
confirmation of the hypothesis of three color degrees of freedom for
quarks.

The current values for R below 5 GeV have experimental uncertainties on
the order of 15\%\cite{2,3,4,5}.  These uncertainties limit the precision
of $\azinv$ and $\am$ and in turn the determination of the Higgs mass
from radiative corrections in the standard model\cite{6,7,8}.  For
example, the contributions to the value of $\azinv$ and the error in
$\azinv$ are shown in Figures~\ref{fig:alphval} and \ref{fig:alpherr}.

The BES collaboration has started a program of R measurements over the
2-5 GeV energy range with the goal of reducing the present uncertainties
by a factor of two or more.

\begin{center}
\begin{minipage}[t]{3.50in}
\begin{figure}[htb]
\epsfysize=2.5in
\centerline{\epsfbox{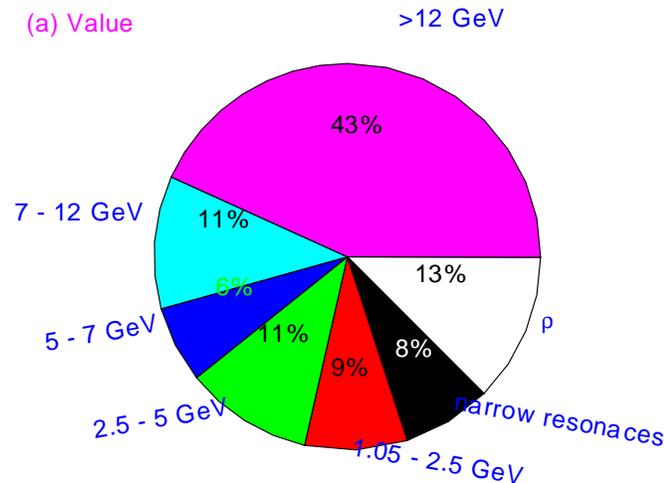}}
\caption{Contribution from different energy ranges to value of $\az$.
Figure from Ref. [9].}
\label{fig:alphval}
\end{figure}
\end{minipage}
\begin{minipage}[t]{3.50in}
\begin{figure}[htb]
\epsfysize=2.5in
\centerline{\epsfbox{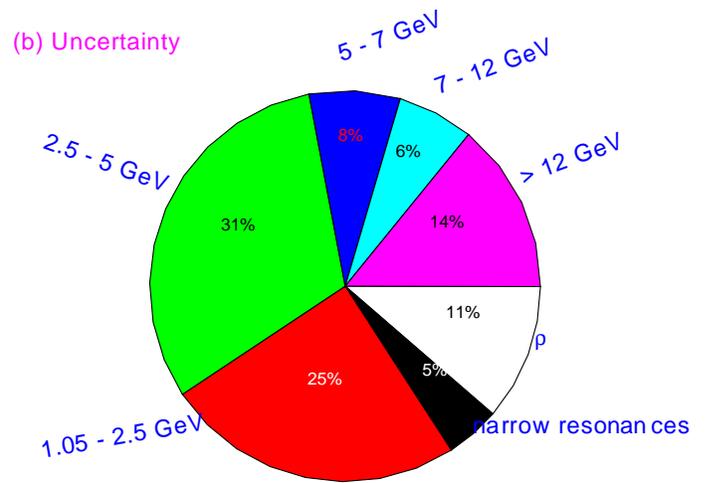}}
\caption{Contribution from different energy ranges to error in $\az$.
Figure from Ref. [9].}
\label{fig:alpherr}
\end{figure}
\end{minipage}
\end{center}

\section{Detector}

The measurements were carried out with the upgraded BES detector
(BESII), which is a conventional solenoidal detector that is described
in some detail in Ref.~\cite{10}.  Upgrades included the replacement of
the central drift chamber with a vertex chamber (VC) composed of 12
tracking layers organized around a beryllium beam pipe.  This chamber
provides a spatial resolution of about 90 $\mu$m.  The barrel
time-of-flight counter (BTOF) was replaced with a new array of 48
plastic scintillators that are read out by fine mesh photomultiplier
tubes situated in the 0.40~T magnetic field volume, providing 180 ps
resolution.  A new main drift chamber (MDC) has 10 superlayers, each
with four sublayers of sense wires.  It provides $dE/dx$ information for
particle identification and has a momentum resolution of $\delta p/p=1.8
\%\sqrt{(1+p^2)}$ for the charged tracks with momentum $p$ in GeV.  The
sampling-type barrel electromagnetic calorimeter (BSC), which covers
80\% of 4$\pi$ solid angle, consists of 560 cells along the $\phi$
direction, with each cell containing 24 layers in the radial direction.
The BSC operates in the self-quenching streamer mode with an energy
resolution of $\delta E/E=21 \%/\sqrt{E}$ (E in GeV) and a spatial
resolution of 7.9 mrad in $\phi$ and 3.6 cm in $z$.  The outermost
component of BESII is a $\mu$ identification system consisting of three
double layers of streamer tubes interspersed in the iron flux return of
the magnet.  These measure coordinates along the muon trajectories with
resolutions of 3 cm and 5.5 cm in $r\phi$ and $z$, respectively.

\section{Data Analysis}

A first scan of $R$ measurements was performed this spring with data
samples of about 1000-2000 hadronic events collected each at 2.6, 3.2,
3.4, 3.55, 4.6 and 5.0 GeV.  To understand beam-associated backgrounds,
separated beam data were taken at each energy point, and single beam
data were accumulated at 3.55 GeV.  Additionally, data were taken at the
$\jp$ and $\psi'$ resonances to provide a precise energy calibration as
well as a cross check of the trigger efficiencies for the Bhabha, dimuon
and hadronic events measured in the R scan.  The 3.4 GeV point was
remeasured after a time interval of about two weeks to test the
stability of the detector.

The value of $R$ is determined from the number of observed hadronic
events ($N^{obs}_{had}$) by the relation

\begin{equation}
R = \frac{\sigma^0(\ee \ra \mbox{hadrons})}{\sigma^0(\ee \ra \mumu)} =
\frac{N^{obs}_{had} - N_{bg} - \sum_{l}N_{ll} - N_{\gamma\gamma}}
{\sigma^0_{\mu\mu} \cdot L \cdot \epsilon_{had} \cdot \epsilon_{trg}
\cdot (1+\delta)},
\end{equation}

\noindent
where $\sigma^{0}$ is the tree-level cross-section of the particular
process, $N_{bg}$ is the number of beam associated background events;
$\sum_{l}N_{ll},~(l=e,\mu,\tau)$ and $N_{\gamma\gamma}$ are the numbers
of misidentified lepton-pair and two-photon processes events; $L$ is the
integrated luminosity; $\delta$ is the correction for initial state
radiation; and $\epsilon_{had}$ and $\epsilon_{trg}$ represents the
detection and trigger efficiency for the hadronic events.  In the
following sections, we examine each of these factors in detail.

\subsection{Run Stability and Performance}

\begin{figure}[htb]
\epsfysize=2.9in
\centerline{\epsfbox{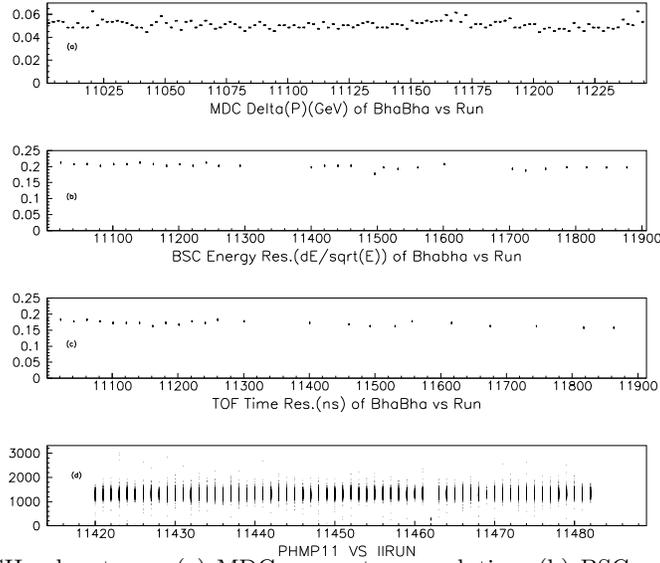}}
\caption{Plots of various BESII subsystems.  (a) MDC momentum
resolution; (b) BSC energy resolution; (c) BTOF time resolution; (d)
MDC dE/dx of the R scan data.  (a)--(c) are based on $\jp$ data.  All
plots are produced using Bhabha events.}
\label{fig:stab}
\end{figure}

A 1.5M $J/\psi$ event sample was used for detector calibration and for
monitoring the data quality and the stability of the detector
performance.  These data indicated that the BESII detector performance
was stable and the data quality was good.  Figure~\ref{fig:stab} shows
the time dependence of MDC momentum resolution, BSC energy resolution,
BTOF time resolution and $dE/dx$ pulse heights for Bhabha events.

\subsection{Hadronic Event Selection}

The goal of hadronic event selection was to distinguish single-photon
hadron production from other processes.  The following track-level
selection criteria were used to define good charged tracks:

\begin{itemize}
\item $|\cos \theta| < 0.84$, where $\theta$ is the track polar angle;
\item The track must have a reasonable three-dimensional helix fit;
\item Distance of closest approach to the beam in the transverse plane
      and along the beam axis are less than 2.0 and 15 cm,
      respectively;
\item $p < p_{beam} + (5 \times \sigma_{p})$, where $p$ and $p_{beam}$
      are the momenta of the track and the beam, respectively, and
      $\sigma_p$ is the momentum resolution of the beam;
\item $E < 0.6 E_{beam}, E/p < 0.8$, where $E$ and $E_{beam}$ are the
      energy of the track (as measured in the BSC) and beam,
      respectively, and $p$ is again the momentum of the track;
\item A track must not be definitely identified as an electron via dE/dx
      information;
\item A track must not be definitely identified as a muon;
\item $2 < t < t_{p} + (5 \times \sigma_{t})$ (in ns), where $t$ and
      $t_{p}$ are the time-of-flight for the track and a nominal
      time-of-flight calculated for the track assuming a proton
      hypothesis, respectively, and $\sigma_{t}$ is the BTOF time
      resolution.
\end{itemize}

After the track-level selection, a further event level selection was
applied via the following:

\begin{itemize}
\item At least 2 good charged tracks, with at least one track having a
      good helix fit;
\item A total deposited energy in the BSC $> 0.28 E_{beam}$.
\end{itemize}

A further selection scheme was required based on the number of good
tracks in the event.  For three or more prong events, the only
additional requirement was that all the charged tracks not be positive
(to remove beam-gas events).  However, two-prong events needed to be
distinguished from cosmic ray and lepton pair events, requiring a
further selection scheme:

\begin{itemize}
\item The two tracks must not have been back-to-back:

\begin{equation}
|\theta_{1} + \theta_{2} - 180^{\circ} | > 12^{\circ}, | |\phi_{1} -
 \phi_{2}| - 180^{\circ} | > 4^{\circ};
\end{equation}

\item At least two isolated neutral tracks fulfilled the following:

\begin{equation}
E_{\gamma} > 60 \mbox{MeV},|\theta_{\gamma} - \theta_{c}| > 15^{\circ},
|\phi_{\gamma} - \phi_{c}| > 30^{\circ},
\end{equation}

\noindent
where $\theta_{c}$ and $\phi_{c}$ are the $\theta$ and $\phi$ angle of
either of the two charged tracks.
\end{itemize}

\begin{figure}[htb]
\epsfysize=2.5in
\centerline{\epsfbox{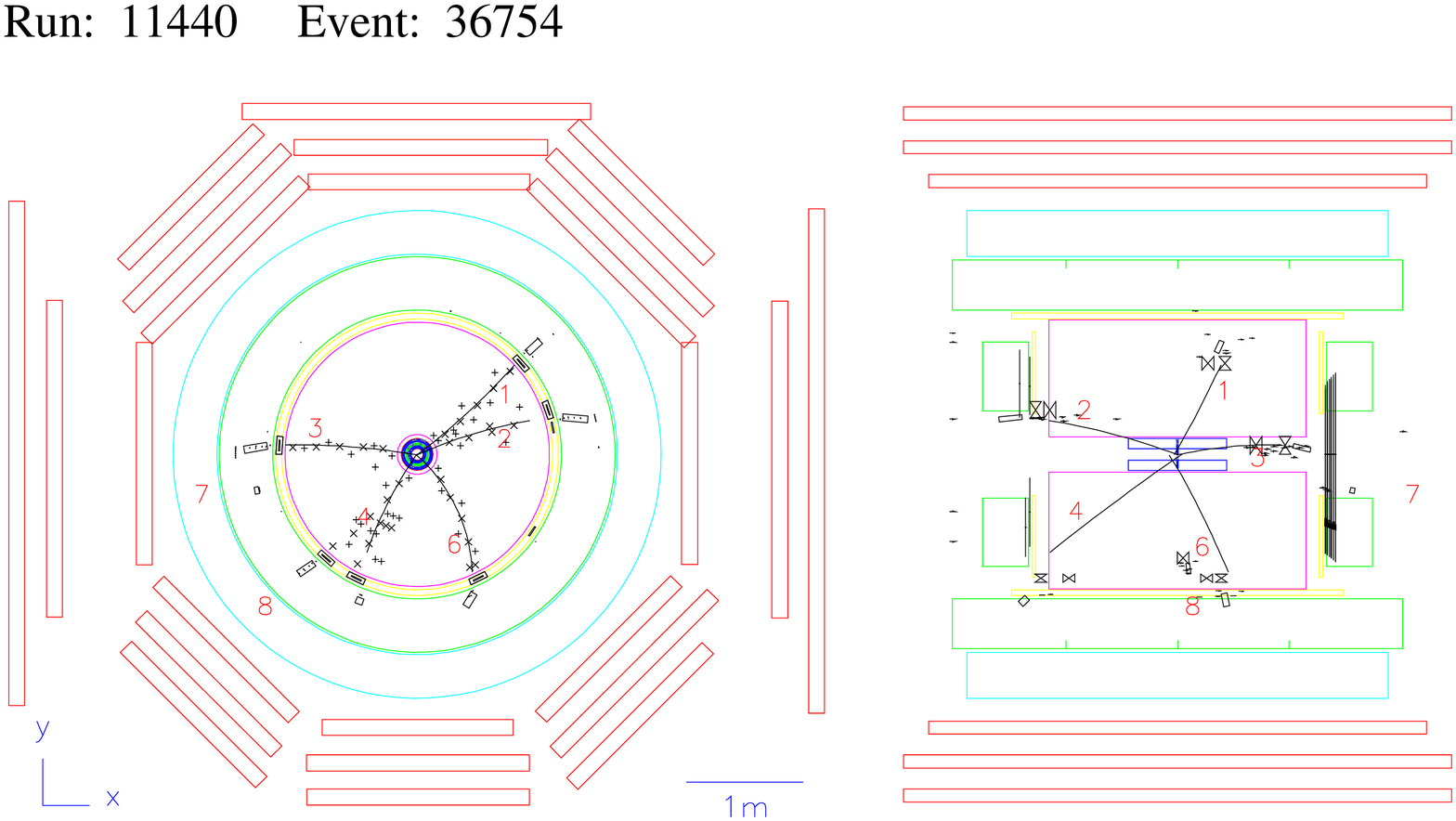}}
\caption{A typical hadronic event display from 3.55 GeV data.}
\label{fig:event}
\end{figure}

The performance of the event selection routines were checked by hand
scans, which were carried out for both the selected and rejected
hadronic events, the separated-beam data, and the Monte Carlo events.  A
typical hadronic event that passed the selection cuts is shown in
figure~\ref{fig:event}.  We determined the error associated with the
hadronic selection criteria from changes in the event yield caused by
varying the selection requirements and from the results of the hand
scan.

\subsection{Background Subtraction}

There were three major types of background to be considered.  One type,
consisting of cosmic rays, Bhabha, dimuon events and some two-photon
process events, was directly selected out during the event selection
routine.  The second, consisting of tau-pair production and residual
two-photon processes, was subtracted out statistically via a Monte Carlo
simulation.

Finally, the most serious sources of background in the hadronic event
sample were beam-gas and beam-wall interactions.  Most of the beam-gas
and beam-wall background events were rejected by a vertex cut.  The
salient features of the beam associated background were that their
tracks were very much along the beam pipe direction, the energy
deposited in BSC was small, and most of the tracks were protons.  The
same hadronic event selection criteria were applied to the
separated-beam data, and the number of separated-beam events $N_{sep}$
surviving these criteria were obtained.  The number of the beam
associated background events $N_{bg}$ in the corresponding hadronic
event sample was given by $N_{bg}=f \times N_{sep}$, where $f$ was
determined by the ratio of the products of the pressure P at the
collision region times the integrated beam currents I, i.e.
$N_{bg}=N_{sep} \cdot (\int\limits_{run} \ P_{run} \cdot I_{run}\,dt)/
(\int\limits_{sep} \ P_{sep} \cdot I_{sep}\,dt)$.

\subsection{Luminosity}

The integrated luminosity was determined using large-angle Bhabha events
with the following selection criteria, using only BSC information:

\begin{itemize}
\item Two clusters in the BSC with largest deposited energy in the
      polar angle $|\cos \theta| \le 0.55$;
\item Each cluster with energy $> 1.0$ GeV (for 3.55 GeV data, scaled
      for other energy points);
\item $2^{\circ} < | |\phi_{1} - \phi{2}| - 180^{\circ} | < 16^{\circ}$,
      where $\phi_{1}$ and $\phi_{2}$ are the azimuthal angles of the
      clusters to account for radiative events.
\end{itemize}

A cross check using only MDC information (dE/dx) was generally
consistent with the BSC measurement; the difference was taken into
account in the systematic error.

\subsection{Hadronic Detection Efficiency}

\begin{figure}[htb]
\epsfysize=2.6in
\centerline{\epsfbox{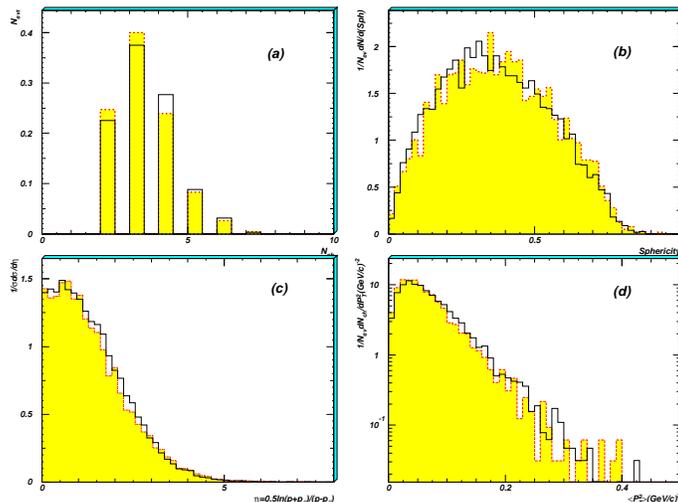}}
\caption{Comparison of hadronic event shapes between data (shaded
region) and Monte Carlo (histogram).  (a) Multiplicity; (b) Sphericity;
(c) Rapidity; (d) Transverse momentum.}
\label{fig:lund}
\end{figure}

The detection efficiency for hadronic events was determined via a Monte
Carlo simulation using the JETSET7.4 event generator\cite{11}.
Parameters in the JETSET7.4 generator were tuned using more than 40,000
hadronic events selected from the tau mass measurement data
sample\cite{12}.  The parameters of the generator were adjusted to
reproduce distributions of kinematic variables such as multiplicity,
sphericity, transverse momentum, etc.  Figure~\ref{fig:lund} shows the
charge multiplicity, the sphericity, the aplanarity and the transverse
momentum distributions for the real and simulated event samples.

\subsection{Trigger}

A two-level trigger system retained good events while rejecting
background~\cite{10}.  The trigger was derived from signals from a beam
pickup electrode located upstream of the detector and from each
sub-detector.  Event categories were classified according to numbers of
charged and neutral tracks seen at the trigger level.  For beam
crossings with charged tracks, two trigger channels were utilized: in
the first, we required at least one hit in the 48 BTOF counter array,
one track in the VC and MDC, and at least 150~MeV of energy deposited in
the BSC; in the second, we required back-to-back hits in the BTOF
counter with one track in the VC and two tracks in the MDC.  For the
neutral track trigger, we required that the sum of the deposited energy
of the tracks in two cells of the BSC was greater than 80 MeV and the
total energy deposited in BSC from all sources was greater than 800
MeV.

All coincidences were formed using 40, 120, 630, 640 ns wide gates
derived from the beam pickup signal for the TOF, VC, MDC and BSC,
respectively.  Event candidates fulfilling the two-level criteria were
accepted.

The trigger efficiencies were measured run by run by comparing the
responses to different trigger requirements.  As a cross check, data
taken in special runs made at the $J/\psi$ resonance was used to provide
independent measurements of the trigger efficiencies.
Table~\ref{tab:trigeff} lists the measured efficiencies for different
event types in the R scan data, as well as the cross-check performed
using $\jp$ data.  The errors in the trigger efficiencies for Bhabha and
hadronic events were less than $\pm$0.5\%.

\begin{table}[htb]
\caption{Trigger efficiencies.}
\begin{tabular}{lrrr}
           & \multicolumn{3}{c}{Event type} \\
Data set   & Bhabha & dimuon & hadronic \\
\tableline
R data     & 99.5\% & 98.2\% &   99.2\% \\
$\jp$ data & 99.6\% & 98.2\% &   99.9\% \\
\end{tabular}
\label{tab:trigeff}
\end{table}

\nopagebreak
\subsection{Initial State Radiation}

The effect of radiative corrections on the measured cross-section was
examined in detail.  Four different schemes were examined: Berends and
Kleiss\cite{13}, Kuraev and Fadin\cite{14}, Bonneau and Martin\cite{15},
and a method used by the Crystal Ball\cite{16}.  Radiative corrections
calculated by all these methods were consistent to within 1\% for
off-resonance points, but could differ by 1-3\% in resonance regions.

We used the approach of Bonneau and Martin, with the differences from
the other schemes included in the systematic error.

\subsection{Preliminary Results}

The preliminary R values obtained at the six energy points are shown in
Table~\ref{tab:rvalue} and graphically displayed in
Figure~\ref{fig:rvalue}.  A further breakdown of contributions to the
systematic errors is given in Table~\ref{tab:rsyst}.

\vspace{-0.05in}
\begin{table}[h]
\caption{Summary of R data and values.}
\begin{tabular}{cccccccccc}
$E_{cm}$ & $N_{had}^{obs}$ & $N_{bg}$ & $\cal{L}$ (nb$^{-1}$) &
$\epsilon_{had}(\%)$ & $(1+\delta)$ & $\sigma^{0}_{\mu\mu}$ & R value &
Systematic error & Statistical error \\
\tableline
2.60 & 5617 &  8 & 292.9 & 0.5482 & 1.009 & 12.840 & 2.61 & 6.59\% & 1.36\% \\
3.20 & 2051 & 10 & 109.3 & 0.6430 & 1.447 &  8.477 & 2.25 & 5.65\% & 2.27\% \\
3.40 & 2149 & 46 & 135.3 & 0.6961 & 1.173 &  7.509 & 2.36 & 6.80\% & 2.26\% \\
3.55 & 2672 & 50 & 200.2 & 0.6840 & 1.125 &  6.888 & 2.29 & 7.17\% & 2.03\% \\
4.60 & 1497 & 48 &  87.7 & 0.8227 & 1.079 &  4.102 & 3.53 & 8.93\% & 2.98\% \\
5.00 & 1648 & 26 & 102.3 & 0.8453 & 1.068 &  3.472 & 3.42 &10.27\% & 3.02\% \\
\end{tabular}
\label{tab:rvalue}
\end{table}

\vspace{-0.1in}
\begin{table}[h]
\caption{Contributions to systematic errors.  All errors are in
percentages (\%).}
\begin{tabular}{ccccccccc}
$E_{cm}$ & Had. select & Background & $\cal{L}$ & $\tau$-pair & Bhabhas
& Had. efficiency & Trigger & Rad. corr \\
\tableline
2.60 & 5.25 & 0.06 & 2.12 & 0.00 & 0.04 & 2.60 & 0.50 & 2.00 \\
3.20 & 4.03 & 0.15 & 2.83 & 0.00 & 0.04 & 1.90 & 0.50 & 2.00 \\
3.40 & 5.06 & 0.27 & 2.83 & 0.00 & 0.04 & 2.90 & 0.50 & 2.00 \\
3.55 & 6.05 & 0.27 & 2.32 & 0.00 & 0.04 & 2.30 & 0.50 & 2.00 \\
4.60 & 7.58 & 0.75 & 2.16 & 0.32 & 0.00 & 3.60 & 0.50 & 2.00 \\
5.00 & 9.00 & 1.26 & 2.81 & 0.32 & 0.00 & 3.20 & 0.50 & 2.00 \\
\end{tabular}
\label{tab:rsyst}
\end{table}

\section{Future Plans}

The main goal of the first scan was a careful understanding of the
trigger and hadronic event acceptances, as well as the hadronic event
selection and the background subtraction, which are central to a total
cross section measurement.  In 1999, BES-II will carry out a finer scan
from 2 to 5 GeV as indicated by Figure~\ref{fig:future}.  Our goal is to
reduce the present uncertainties of $R$ by at least a factor of two over
this entire energy region.

\nopagebreak
\begin{minipage}[t]{4.00in}
\begin{figure}[htb]
\epsfysize=2.5in
\centerline{\epsfbox{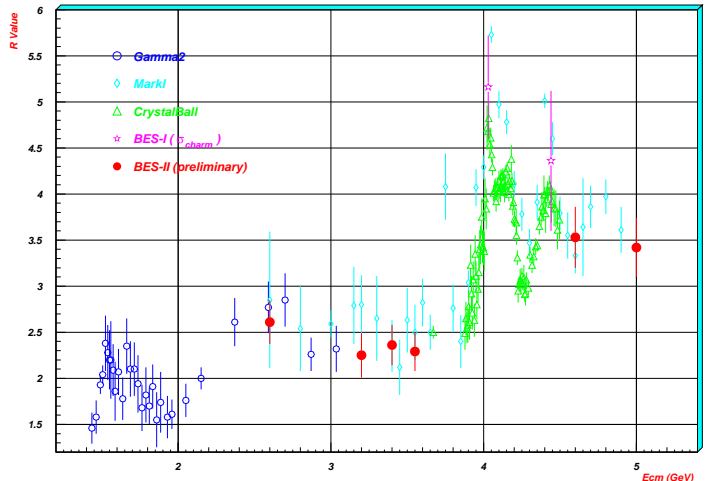}}
\vskip -.1 cm
\caption{Plot of R values at the 6 data points vs $E_{cm}$.}
\label{fig:rvalue}
\end{figure}
\end{minipage}
\begin{minipage}[t]{2.75in}
\begin{figure}[htb]
\epsfysize=2.5in
\centerline{\epsfbox{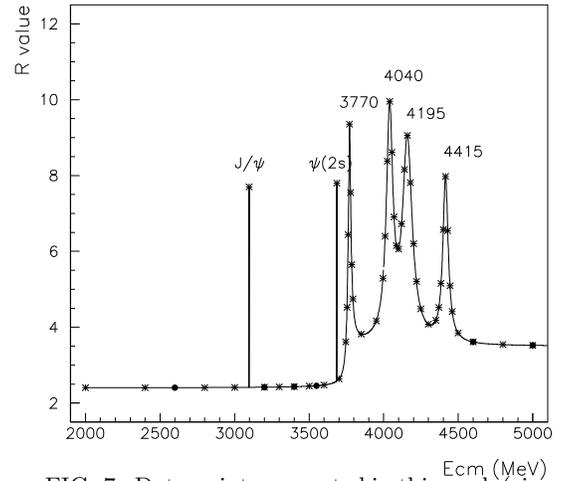}}
\vskip -.1 cm
\caption{Data points presented in this work (circles) and points to be
scanned during the Spring 1999 run.  The R curve shown is a
representation of the resonance region and does not reflect actual
measurements.}
\label{fig:future}
\end{figure}
\end{minipage}


\section{References}

\begin{references}

\bibitem{1} Particle Data Group, Eur. Phys. J. {\bf C3} 204 (1998).
\bibitem{2} F. Ceradini et al., Phys. Lett. {\bf B47} (1973) 80;\\
            B. Batoli et al., Phys. Rev. {\bf D6} (1972) 2374;\\
            M. Bernardini et al., Phys. Lett. {\bf B51} (1974) 200.
\bibitem{3} G. Cosme et al., Phys. Lett. {\bf B40} (1972) 685.
\bibitem{4} M. Kurdadze et al., Phys. Lett. {\bf B42} (1972) 515.
\bibitem{5} A. Litke et al., Phys. Rev. Lett. {\bf 30} (1973) 1189.
\bibitem{6} A. Blondel, plenary talk at ICHEP'96, Warsaw.
\bibitem{7} B. Pietrzyk, Rad. Corr. Conf., Cracow 1996.
\bibitem{8} M. Davier and A. Hoeker, LAL 97-85.
\bibitem{9} H. Burkhardt and B. Pietrzyk, Phys. Lett. {\bf B356} (1995)
            398.
\bibitem{10} J.Z. Bai et al., Nucl. Instr. and Methods in Phys. Research
             {\bf A344} (1994) 319.
\bibitem{11} Torbjoen Sjoestrand, LU TP 95-20.
\bibitem{12} X.R. Qi et al., to be published on High Energy Physics and
             Nuclear Physics (in Chinese).
\bibitem{13} F.A. Berends and R. Kleiss, Nucl. Phys. {\bf B178} (1981)
             141. 
\bibitem{14} E. A. Kuraev and V. S. Fadin, Sov. J. Nucl. Phys. {\bf 41}
             (3) (1985) 3.
\bibitem{15} G. Bonneau and F. Martin, Nucl. Phys. {\bf B27} (1971)
             387. 
\bibitem{16} C. Edwards et al., SLAC-PUB 5160 (1990)
\end{references}
\end{document}